\documentclass[12pt]{iopart}

\usepackage{amssymb}
\usepackage{iopams}
\usepackage[dvips]{graphicx}
\usepackage{chemarrow}
\usepackage{mathrsfs}
\usepackage[all]{xy}
\usepackage[ps2pdf,colorlinks=true,pdfpagemode=UseNone, urlcolor=blue,linkcolor=blue,citecolor=red]{hyperref}
\usepackage{url}

\newtheorem{theorem}{Theorem}[section]
\newtheorem{corollary}[theorem]{Corollary}
\newtheorem{lem}[theorem]{Lemma}
\newtheorem{proposition}[theorem]{Proposition}

\newtheorem{definition}{Definition}[section]
\newtheorem{remark}{Remark}[section]

\begin{document}

\title[Symplectic Regularization of the Circular $N$+2 Sitnikov Problem]{}
{Symplectic Regularization of Binary Collisions in the Circular $N$+2 Sitnikov
Problem}
\author{Hugo Jim\'enez-P\'erez}
\address{IMCCE, Observatoire de Paris,\\
77 av. Denfert-Rochereau,\\ 
75014 Paris, France}
\ead{\mailto{jimenez@imcce.fr}}
\author{Ernesto A. Lacomba}
\address{Dept. of Math., Univ. Aut. Metrop.,\\ San Rafael Atlixco 186,
C.P. 09430,\\ Iztapalapa, Mexico City}
\ead{\mailto{lace@xanum.uam.mx}}

\submitto{\JPA}

\begin{abstract}
We present a brief overview of the regularizing transformations of the Kepler problem
and we relate the Euler transformation with the symplectic structure of the phase space of the $N$-body
problem. We show that any particular solution of the $N$-body problem
where  two bodies have rectilinear dynamics can be regularized
by a linear symplectic transformation and the inclusion of the Euler transformation
into the group of symplectic local diffeomorphisms over the phase space.
As an application we  regularize a particular configuration of the
restricted circular $N$+2 body problem.
\end{abstract}

\maketitle

\section*{Introduction}
In Celestial Mechanics, the $N$-body problem has two types of singularities:
collisions between two or more bodies, and escapes in bounded  time.
In order to study the behavior of the system
close to singularities, it is a common procedure to transform it to another
equivalent system that avoids the singularities by means of some methods
called \emph{regularizations}. There are a lot of regularizing transformations,
but unfortunately it is not always possible to regularize an arbitrary singularity.
For instance, infinite expansions at finite time produce esential singularities in the
mathematical model that are not
regularizable by topological or analytical methods known until now,
and the same is true for some multiple collisions.

Basically there exist two types of regularizations: analytic regularization
formalized by Siegel and Moser \cite{Sie1}, regularization by surgery or 
topological (also known as block regularization) discovered by
Conley and Easton \cite{Con1, Eas1}. In particular, it is
well-known that collisions between two
infinitesimal bodies (in $N$+$\nu$ problems) and triple collisions are
impossible to regularize by the Easton method \cite{Mar1}. Marchal \cite{Mar1} has a
very clear exposition about the classification of the singularities in the
$N$-body problem and their regularization (when it is possible).

In this paper we deal with the analytical or Siegel's regularization
\cite{Sie1} which is achieved by three ingredients: a local change
of coordinates by means of some local diffeomorphism $\rho:M\to M$
on the phase space, a scaling function $g:M\to \mathbb R$ that
introduce a new fictitious time $\tau$ by the relation
$\frac{dt}{d\tau}=g(w)$ and a set of initial conditions
$\phi_0=\phi(0)$ of the flow which specifies the solutions that go
to collision; since  $N$-body problems are  Hamiltonian problems,
the set of initial conditions determines an energy level by
the conservation of the energy. It means that in Hamiltonian
problems the analytical regularization process is performed on each
fixed energy level $H(x)=h$.

  Thus, the process is as follows:
\begin{itemize}
 \item choose a fixed energy level $H(x)=h$ and consider
$$
(H-h)(x)=0,
$$
 \item apply the change of coordinates $x=\rho(w)$ of the phase space
$$
[(H-h)\circ \rho](w)=0
$$
  \item apply the scaling transformation $\frac{dt}{d\tau}=g(w)$ multiplying
the last expression
$$
 \Gamma=[g\cdot(H-h)\circ \rho](w)
$$
  \item the preimage of $\Gamma(w,h)=0$ generates the energy levels of the  regularized
system for each $h\in {\rm Img}(H)\subset\mathbb R$ fixed.
\end{itemize}
    It is important to keep in mind that the aim of regularization theory is
to transform singular differential equations into regular ones, controlling the
velocity of the regularized system by the scaling time \cite{Cel1}.

For the one-dimensional Kepler motion with Hamiltonian function
\begin{eqnarray}
 H(q,p)= \frac12 y^2 -\frac{\mu}{x},\qquad x\in \mathbb R^*,\quad y\in T_x\mathbb R,\label{eqn:kep1}
\end{eqnarray}
it was already found by Euler that the
introduction of a square-root coordinate $u=\sqrt{x}$  and a fictitious time $\tau$ defined by
$dt=xd\tau$ reduces the Kepler equation of motion (\ref{eqn:kep1}) to
the equation of motion of a one-dimensional harmonic oscillator
\begin{eqnarray}
 \mu = \frac12 v^2 + h\ u^2 ,\qquad u\in \mathbb R^*, \quad v\in T_u\mathbb R,\label{eqn:reg:kep1}
\end{eqnarray}
if $h < 0$ \cite{Bar1}; where $y=\frac{dx}{dt}$, $v=\frac{du}{d\tau}$ and $\mathbb R^*=(\mathbb R - \{0\})$.

Generalizing this approach,   Levi-Civita introduces its
``\emph{...transformation du syst\`eme qui donne lieu \`a des
cons\'equences remarquables...}'' in \cite[1907]{Lev1}. In his work
Levi-Civita introduces a conformal transformation and
exploits the symplectic structure of the complex plane $(\mathbb
C,dz\wedge d\bar{z})\cong(\mathbb R^2,dy\wedge dx)$. In fact, this
regularization is made on the cotangent bundle $T^*(\mathbb{C}^*)$
where $\mathbb {C}^*=\{z\in\mathbb C: z\neq 0\}$ is viewed as an
open symplectic manifold. Levi-Civita regularization is achieved by
the local diffeomorphism
\begin{eqnarray}
\left.
\begin{array}{rcl}
    \rho:(T^*(\mathbb{C}^*),\omega)&\to&(T^*(\mathbb C^*),\omega) \\
              (z,w) &\mapsto& \left(z^2, \frac{w}{2|z|}\right)
\end{array}
\right.\label{eqn:levi}
\end{eqnarray}
and the time rescaling  $dt=|z|^2d\tau$.
The transformation above takes the Hamiltonian function
\begin{eqnarray}
 H(q,p)= \frac12 p^2 -\frac{\mu}{|q|},\qquad q\in \mathbb C^*, \quad p\in T_q\mathbb C,
\end{eqnarray}
into the equation
\begin{eqnarray}
 \mu = \frac12 w^2 + h|z|^2,\qquad z\in \mathbb C^*,\quad w\in T_z\mathbb C,
\end{eqnarray}
where $p=\frac{dq}{dt}$,
 $w=\frac{dz}{d\tau}$ and the symplectic form in the regularized phase space is
$\omega=d\bar w\wedge dz$. The expression (\ref{eqn:levi}) is a contact transformation since
it preserves the canonical Liouville 1-form $\alpha=\bar w dz$. If we denote
the image of the local diffeomorphism by
$(q,p)=\rho(z,w)$ then
\begin{eqnarray}
   \bar{p}dq &=& \bar{w}dz,\label{eqn:1form}
\end{eqnarray}
that is, $\rho^*(\alpha)=\alpha$; as a consequence we have a
symplectic (canonical) transformation. Applying the exterior
differential to both sides of (\ref{eqn:1form}) we obtain the
symplecticity condition $\rho^*(\omega)=\omega$ for the
transformation. In 1913 Sundman introduced a transformation that
maps the unitary circle in $\mathbb R^2$  into the band
$-1<y<1$, and obviously this mapping does not preserves the area
\cite[pp 127-129]{Hag1}.

Unfortunately, the procedure described above is difficult to
generalize to the 3-dimensional case since the euclidean
space $\mathbb R^3$ does not posses any complex structure.
However,  the Kustaanheimo-Stiefel's regularization \cite{Kus1},
generalizes the Levi-Civita regularization to the four dimensional
complex manifold $T^*\mathbb C^2$ (real dimension 8) and projects it
onto some symplectic submanifold of real dimension 6 \cite{Sti1}. In recent
years, the K-S transformation using quaternions and the quaternionic
algebra has gained much attention,
 from the works of Vivarelli \cite{Viv1},  Volk \cite{Vol1}, Vrbik \cite{Vrb1},
Waldvogel \cite{Wal1,Wal2}, among others

On the other hand, some of the most recent works for computing collision 
orbits using symplectic integrators are based on the \emph{algorithmic 
regularization}.
This procedure was introduced by Mikkola and Tanikawa \cite{Mik1,Mik2}
simultaneously with Preto and Tremaine \cite{Pre1} in 1999. Algorithmic 
regularization uses a particular time scaling function 
\begin{eqnarray*}
  \frac{dt}{d\tau} =g(w, t) ,\quad (w,t)\in T^*\mathcal Q\times \mathbb R,
\end{eqnarray*}
defined 
on the extended phase space, instead of the classical
$g(q)=f(q)\prod_{i,j}r_{ij}$, where $q\in\mathcal Q$ and $r_{ij}=\sqrt{q_i-q_j}$.
The more interesting property of algorithmic regularizations 
is the absence of a coordinate transformation. 

In order to construct the time scaling function $g(w,t)$, the extended phase 
space $T^*\mathcal Q\times \mathbb R$ is considered as a presymplectic manifold
and then immersed into a symplectic one, locally diffeomorphic to
$(\hat M, \omega_{\hat M})$ where
$\hat M=(T^*\mathcal Q\times T^*\mathbb R)$ and $\omega_{\hat M}
=\omega -dt\wedge dH$\footnote{In fact, algorithmic regularizations are 
selected by their numerical properties and the separability of the regularized 
system, in order to facilitate the numerical computations with symplectic 
integrators like the \emph{leapfrog} scheme.}. Then, we search for a 
function $g:\hat M\to\mathbb R$
such that the resulting Hamiltonian function $\Lambda = g(z)(H(q,p,t)-h)$
will be separable.

At this point, there exists two types of algorithmic regularization: the
logarithmic Hamiltonian and the Time Transformed Leapfrog (TTL). The former
is a canonical extension of the original Hamiltonian system to the extended
symplectic manifold $(\hat M,\omega_{\hat M})$. The Hamiltonian function
$H(q,p)=T(p)-V(q)$ extends to the function 
\begin{eqnarray*}
\Lambda(Q,P) = \log\left(T_e(P)\right)-\log(V(Q))\quad 
\end{eqnarray*}
where $P=(p,h)$, $Q=(q,t)$, $T_e(P)=T(p)-h$ and $H=h$ is a fixed value. 
The new independent variable is $\tau =\int_0^t T(p)-h\ ds$
and the Hamiltonian vector field $X_{\Lambda}$ becomes 
$\dot z=J\nabla_z \Lambda(z)$, $z=(Q,P)$.

TTL is a non-canonical generalization of the logarithmic
Hamiltonian. In this case, the scaling function $g$ contains the term
$\Omega=\sum_{i<j}(\Omega_{ij}/r_{ij})$ for some selected coefficients
$\Omega_{ij}\in\mathbb R$. The vector field 
\begin{eqnarray*}
\dot q = A^{-1}p, \qquad \dot p=F(q) 
\end{eqnarray*}
is transformed into 
\begin{eqnarray*}
  q^\prime = A^{-1}p/W,\qquad t^\prime = 1/W, \qquad 
  p^\prime = F(q)/W, \qquad W^\prime =\frac{\partial \Omega}{\partial q}p.
\end{eqnarray*}
and regularization of two body collisions is obtained if $\Omega \sim 1/r$ 
near collisions. 
These ``regularizations'' have shown a satisfactory behavior in
numerical computations close to collisions. However, their geometrical 
analysis will be considered by the authors in a future work.

\section{Symplectic Structure of Regularizing Transformations}

In symplectic geometry, mechanical problems are represented by
Hamiltonian systems $(M,\omega,X_H)$ on the phase space viewed as a symplectic
manifold. The standard symplectic manifold is the cotangent bundle
$M = T^*\mathcal Q$ of the configuration space $\mathcal Q = (\mathbb R^{N\cdot n}-\Delta)$,
 where $\Delta$ is the set of the singularities of $X_H$ and $H$. This manifold is provided with
the canonical symplectic form $\omega=d{\bf p}\wedge d{\bf q}$ where
${\bf q}\in\mathcal Q$ and ${\bf p}\in T^*_{\bf q}\mathcal Q$.

In particular, problems on celestial mechanics are based on the
Newtonian $N$-body equations,
$
   \mathcal M \ddot{\bf q} = -\frac{\partial V}{\partial {\bf q}}
$
where
\begin{eqnarray}
   V({\bf q}) = G\sum_{i,j}^N \frac{m_im_j}{|q_i-q_j|}, \quad {\bf q}=(q_1,\dots,q_N),
    \quad q_i\in \mathbb R^n, \quad 1\leq i\leq N,
\end{eqnarray}
$q_i$ is the position of the $i$-th body, $m_i$ its mass and
$\mathcal M={\rm diag}(m_1I_n,\dots, m_NI_n)$.
Depending on the value of $n$, we refer to this problem as the rectilinear or
collinear problem if $n=1$, the planar problem if $n=2$ and the spatial
problem if $n=3$.
In its Hamiltonian formulation the Hamiltonian function 
$H:M\to \mathbb R$ is defined by
\begin{eqnarray}
   H({\bf q},{\bf p}) = T({\bf p}) - V({\bf q})
\end{eqnarray}
where $T({\bf p})= {\bf p}^T\mathcal M^{-1}{\bf p}$ is the kinetic energy.

It is  clear that the set of singularities  comes
from the potential function $V({\bf q})$. As we have said, it is
not always possible to regularize any arbitrary singularity, however
in this paper we are concerned with sigularities due to binary
rectilinear collisions as the generalization of the rectilinear
Kepler problem. To avoid this type of singularities we perform a
regularizing transformation using a local diffeomorphism $f:M\to M$
 and a time rescaling $g:M\to\mathbb R$.
In some specific cases when it is desirable to preserve the fibers
and sections of the cotangent bundle, the diffeomorphism and the
time rescaling are applied to the base space $f:\mathcal
Q\to\mathcal Q$ and $g:\mathcal Q\to\mathbb R$. To obtain a
local symplectic diffeomorphism on $M=T^*\mathcal Q$  one uses the
properties of the cotangent lift of $f$.

\begin{definition} Let $\mathcal Q$ be an arbitrary differentiable
manifold with cotangent bundle $M=T^*\mathcal Q$, and let  $f\in
{\rm Diff}_x(\mathcal Q)$ be any local diffeomorphism over $\mathcal Q$,
we define the \emph{cotangent lift} of $f$ by
\begin{eqnarray}
        F:=T^*f : M \to M ;\hspace{30pt}     F (p_1) = p_2,
\end{eqnarray}
where $p_i=(x_i,\xi_i)\in M$, $i=1,2$ and $x_i\in \mathcal Q$, $\xi_i \in T^*_{x_i}\mathcal Q$,
and $(df_{x_1})^*\xi_2 =\xi_1$,
\end{definition}

Adittionaly we can see that
\[
   (df_{x_1})^*:T^*_{x_2}\mathcal Q \to T^*_{x_1}\mathcal Q,
\]
so the restriction of $F | _{T^*_{x_1}}$ is the inverse mapping of  $(df_{x_1})^*$.

\begin{proposition}
   The cotangent lift $F$ of any local diffeomorphism $f\in {\rm Diff}_x
   (\mathcal Q)$ is
a local symplectomorphism, which means that
\begin{eqnarray*}
    F^*\omega=\omega.
\end{eqnarray*}
where $\omega$ is the canonical symplectic form
on $M$.
\end{proposition}
The standard references where the reader
can check the proof are \cite[pp 487]{Abr2} and \cite[pp 180]{Abr1}.
It is easy to show that the mapping
\begin{eqnarray*}
  T^* :{\rm Diff}_x(\mathcal Q)& \to& {\rm Sp}_{(x,\xi)}(M,\omega)\\
        \qquad \qquad f &\mapsto& F:=T^*f.
\end{eqnarray*}
is a homomorphism of groups. (Hereafter all concerned diffeomorphisms
are local diffeomorphisms.)

In this way, it is possible to construct symplectomorphisms
that preserve the structure of the cotangent bundle in the sense that
they are \emph{fiberwise} transformations. They
form a subgroup of ${\rm Sp}(M)$ closely related to the
set of generating functions on $M$.

For $N$-body problems in the plane it is a common procedure to
identify the real plane with the complex numbers $\mathbb R^2\cong \mathbb C$.
Szebehely \cite{Sze1} has noted that in order to have a suitable regularizing
transformation for binary collisions in the restricted plane 3-body problem,
the conditions
\begin{eqnarray}
     z = f (w)\qquad{\rm and }\qquad   \frac{dt}{d\tau}= |f_w(w)|^2,  \label{eqn:cond}
\end{eqnarray}
must hold, where $f:\mathbb C\to\mathbb C$ is a meromorphic function of the complex variable $w = u + iv$.

The expression (\ref{eqn:cond}) is a \emph{fiberwise} transformation which preserves the
cotangent bundle as  consequence of the cotangent lift of
$f:\mathbb C\to \mathbb C$ to $T^*\mathbb C$. In such a case, the
bilinear form $\omega_{\mathbb C}=d\bar {p_z}\wedge dz$ gives the symplectic stucture to
$T^*\mathbb C$. Moreover, any
fiberwise symplectic regularization of binary collisions in the $N$ center problem
 has the form equivalent to (\ref{eqn:cond}) \cite{Jim5}.
This condition can be generalized
for symplectic regularizations in higher dimensional spaces
as it is exposed in \cite{Jim1} .

As we have said, it  was known by Euler that the
transformation $u=\sqrt{x}$ and the time rescaling $dt=xd\tau$
reduces the one-dimensional Kepler problem to the one-dimensional
harmonic oscillator for $h<0$. This transformation can be rewritten
as $x=u^2 /2$ with time rescaling $dt=u^2 d\tau$ and it
fulfills condition (\ref{eqn:cond}) when we restrict $x>0$ and
$f:\mathbb R^*\to \mathbb R$.

In order to simplify calculations  and preserve the
symplectic structure we plug in the coefficient $1/2$ to
the transformation and considering the cotangent lift we obtain
\begin{eqnarray}
  x=\frac{u^2}{2},\qquad y=\frac vu, \qquad dt=u^2 d\tau.
\end{eqnarray}
where $y=\frac{dx}{dt}$ and $v=\frac{du}{d\tau}$.
In what follows, we rename the variables $x=q$, $u=Q$, $y=p$ and
$v=P$ to agree  with the standard notation of Hamiltonian
mechanics.
\begin{definition}
   Let $\mathcal N=\{x\in\mathbb R:x>0\}$ be the positive open ray and let $\mathcal V=T^*\mathcal N$ be
its cotangent bundle. We define de \emph{Euler transformation}
$\xi:\mathcal V\to \mathcal V$
as the mapping
\begin{eqnarray}
  \xi:(Q,P)\mapsto \left(\frac{Q^2}{2},\frac PQ\right) \label{eqn:euler}
\end{eqnarray}
where $Q\in \mathcal V$ and $P\in T^*_{Q}\mathcal V$.
\end{definition}

We restrict the domain of the Euler transformation to be an open manifold with
boundary, in order to consider this transformation as a local diffemorphism.

\begin{lem}
   The Euler transformation $\xi:\mathcal V\to \mathcal V$ defined in
   (\ref{eqn:euler}) is a (local) symplectomorphism.
\end{lem}

{\it Proof.}
   We obtain the result in a straightforward way since  the Jacobian matrix
\begin{eqnarray}
 (d\xi)=\left(
    \begin{array}{cc}
      Q  &  -\frac {P}{Q^2} \\
    0  &   \frac 1Q
 \end{array}
 \right)
\end{eqnarray}
 is symplectic.
$\hfill\square$

\begin{definition}\label{def:N}
We call   \emph{Euler regularization} of the collinear Kepler
problem to the Euler transformation together with the rescaling
function $dt=Q^2 d\tau$ applied to the equation of movement of the
Kepler problem.
\end{definition}

It is possible to consider the inclusion of the Euler transformation into the group
${\rm Diff}_x(M)$ of local diffeomorphisms of any symplectic manifold $(M,\omega)$
containning  a two-dimensional linear symplectic subspace $\mathcal V$ such
that  $M\cong \mathcal V
\oplus\mathcal V^\omega$.

We recall that a subspace $V\subset E$ of some symplectic vector space
$(E,\omega)$  of dimension $2n$, is called \emph{symplectic} if the
restriction of the symplectic form $\omega|_V$ is injective (non degenerate).

A well-known result about symplectic vector spaces
that will be useful to understand the regularizing transformation
applied to the circular $N$+2 Sitnikov  problem is the following.

\begin{lem}\label{lem:sum}
   Let $(E,\omega)$ be a symplectic vector space and let $V\subset E$ be a linear
subspace. Then $V$ is a symplectic subspace if and only if
\begin{eqnarray}
    E=V\oplus V^\omega
\end{eqnarray}
where $V^\omega$ is the orthogonal subspace to $V$ with respect to the
bilinear form $\omega$. Moreover, $V^\omega$ is a symplectic subspace.
\end{lem}

The proof of this result is found in any book on symplectic geometry.
Now, we procede to construct the regularizing transformation that we
will apply to some symmetric $N$+2 body problems in
the simpler cases: regularization of binary rectilinear collisions of the
infinitesimals.

\index{transformaci\'on de Euler!inclusi\'on can\'onica de la}
\begin{definition}
   The \emph{canonical inclusion} of the Euler transformation into the group ${\rm Diff}_x(M)$
of an open symplectic manifold with boundary $(M,\omega)$ is the local diffeomorphism
\begin{eqnarray}
   i_\xi: \mathcal V\oplus \mathcal V^\omega\to M,
\end{eqnarray}
such that
\begin{eqnarray}
    i_\xi|_{\mathcal V}=\xi \qquad and\qquad
    i_\xi|_{\mathcal V^\omega}=id_{\mathcal V^\omega}.
\end{eqnarray}
\end{definition}

We have the relation $\imath\circ \xi=i_\xi\circ \imath$, therefore the following diagram comutes
\begin{eqnarray*}
\xymatrix{
   M \ar[r]^{i_\xi}& M  \\
    T^*\mathcal N \ar[u]^{\imath}\ar[r]^{\xi}&\ar[u]^{\imath} T^*\mathcal N}
\end{eqnarray*}
where $\imath(T^*\mathcal N)\cong \mathcal V$ is
as in Definition
\ref{def:N}.

\begin{lem}
    The canonical inclusion is a local symplectomorphism
    $i_\xi\in {\rm Sp}_{(x,\xi)}(U,\omega)$ for $U\subset M$  any open subset.
\end{lem}
{\it Proof.}
    This fact is obtained straightforward from de direct sum $\mathcal V\oplus \mathcal V^\omega$,
then the Jacobian matrix of the differential is
\begin{eqnarray}
  d(i_\xi) = \left(
      \begin{array}{c|c}
         (d\xi) & 0 \\
    \hline
      0  &  I_{2(n-1)}
      \end{array}
  \right),
\end{eqnarray}
where $d\xi\in \mathcal M_{2\times 2}$ is the Jacobian matrix of the Euler transformation and
$I_{2(n-1)}$ is the identity matrix in $\mathcal M_{2(n-1)\times 2(n-1)}$.
\hfill$\square$

\section{Some symmetric N+2 body problems}
Now, we must to characterize particular solutions of the $N$ body problem
where the Euler regularization is applied in a natural way. Since Euler regularization
only considers the unidimensional (rectilinear) evolution of the colliding bodies
we focus our attention to systems with $N$ massive and 2 infinitesimal bodies and
we call them $N+2$ body problems.

\begin{definition}
    We say that a solution $\varphi(t)=\varphi(\varphi_0,t)$ of the spatial $N$ body problem has
    an $\mathscr R$-\emph{symmetry} around the line $\mathcal L\subset \mathbb R^3$
    if   $\mathscr{R} \in SO(3)$ satisfies the following  properties:
    \begin{itemize}
     \item
    for every $t\in (\alpha,\beta)$ and every state 
    $S=( (q_{i}(t),p_i(t)),\cdots, (q_{N}(t), p_N(t) ))$ the action of
    $\mathscr R$ on $S$ is a cyclic permutation of order $r>1$,
    \item for every $x\in \mathcal L$ we have $\mathscr Rx=x$.
    \end{itemize}
\end{definition}
It is clear that the $\mathscr R$-symmetry applies to the whole
phase space since this is valid in the configuration space for every
$t\in(\alpha,\beta)$. This is equivalent to a selection  of
$\mathscr R$-symmetric initial condition $\varphi(0)=\varphi_0$ in
the phase space $M=T^*(\mathbb R^{3N}-\Delta)$ and follow the flow
$\Phi^t(\varphi_0)$.

\begin{remark}
  It is possible to have the limits $\alpha=-\infty$ or $\beta=\infty$ however, 
  in the general case, the $\mathscr R$-symmetry is valid for solutions which 
  comes or goes to singularities when $t^+ \to \alpha$ or $t ^-\to\beta$.
\end{remark}

\begin{proposition}\label{prop:N+1}
  Let $\varphi(t)$ be an $\mathscr R$-symmetric solution of the spatial $N$ body problem for $t\in (\alpha,\beta)$,
  around the fixed line $\mathcal L\subset\mathbb R^3$ and
  we consider the restricted $N+1$ body problem attaching an infinitesimal 
  body to the $\mathscr R$-symmetric solution.
  If the restricted body has initial conditions
  \begin{eqnarray}
     \left(
      \begin{array}{c}
         q_{I}(t_0)\\
         p_{I}(t_0)
      \end{array}
      \right)=
      \left(
      \begin{array}{c}
         q_{I}^0\\
         p_{I}^0
      \end{array}
      \right), \qquad q_{I}^0,p_{I}^0\in\mathbb R^3,\quad t_0\in (\alpha,\beta)
  \end{eqnarray}
  such that $q_{I}^0\in \mathcal L$ and $q_{I}^0\wedge p_{I}^0=0$,
  then the infinitesimal evolves in rectilinear motion on the line $\mathcal L$ for
  $t\in (a,b)\subset (\alpha,\beta)$.
\end{proposition}
{\it Proof.} Since we are concerned with the evolution of the
 infinitesimal body with position $q_{I}^0\in \mathcal L$,
it is sufficient to show that $\dot p_I(t)$ is parallel to $\mathcal
L$ for $t\in(\alpha,\beta)$.

By hypothesis $\varphi(t)$ is a regular $\mathscr R$-symmetric
solution of the $N$ primary bodies around the line $\mathcal L$ for
$t\in(\alpha, \beta)$. Without lost of generality, we can assume
that  the center of mass of the system is fixed at the
origin and $\mathcal L=(0,0,\tau)$, $\tau\in\mathbb R$ is the
vertical line in the 3 dimensional physical space. We supose also
that the constant of universal gravity is $G=1$. The $\mathscr
R$-symmetry  implies that there exists a natural $r>1$ such that
$r|N$ and $\mathscr R^r=Id$; then $\mathscr R$ is a fixed matrix in
$SO(3)$ with components
\begin{eqnarray}
    \mathscr R=
      \left(
      \begin{array}[h]{ccc}
          \cos \frac{2\pi}{r} & \sin \frac{2\pi}{r} & 0\\
          -\sin \frac{2\pi}{r} & \cos \frac{2\pi}{r} & 0\\
          0 & 0 & 1
      \end{array}
      \right).
    \label{eqn:posR}
\end{eqnarray}

Let $s\in\mathbb N$
be the number of equivalent subsystems of the $N$ body problem under the $\mathscr R$ symmetry so $N=rs$.
 We can decompose the $N$ body system  in $s$ partial subsystems with $r$  bodies each one,
 in rearranging the subindices in the way
 \begin{eqnarray}
   (1,2,\dots , N) \to (1_1,\ \cdots\ ,1_s,2_1,\ \cdots\ ,2_s,\ \cdots\ ,r_1,\ \cdots\ , r_s) \label{eqn:ind}
 \end{eqnarray}
 such that
 \begin{eqnarray}
   \hat m_k:=m_{k_1}=m_{k_2}=\dots=m_{k_r}\label{eqn:mk}
 \end{eqnarray}
 and
 \begin{eqnarray}
   q_{k_j}=\mathscr R q_{k_{j-1}} = \mathscr R^{j-1} q_{k_1},
   \label{eqn:qkj}
 \end{eqnarray}
for $k=1,\dots, s$ and $j=2,\cdots,r$.
Positions of each subsystem can be written as
\begin{eqnarray}
    (q_{k_1},q_{k_2},\cdots, q_{k_r}) &=& (q_{k_1},\mathscr R q_{k_1},\cdots \mathscr R^{r-1} q_{k_1}),\quad k=1,\cdots, s,
    \label{eqn:posq}
\end{eqnarray}
where $ q_{k_j} = (x_{k_j},y_{k_j},z_{k_j})$.

The Hamiltonian  vector field for the infinitesimal body  is
\begin{eqnarray}
  \dot q_I &=& \frac{1}{m_I}p_I\\
  \dot p_I &=& -\sum_{1\le i\le N}\frac{m_Im_i(q_I-q_i)}{|q_I-q_i|^3}.
  \label{eqn:champ}
\end{eqnarray}
Rewritting equation (\ref{eqn:champ}) with reindexing (\ref{eqn:ind}) and expressions (\ref{eqn:mk}) and (\ref{eqn:qkj}) we have
\begin{eqnarray}
  \dot p_I
  &=& -m_I\sum_{1\le k \le s} \left(\hat m_k \sum_{1\le i\le r} \frac{(q_I-\mathscr R^i q_{k_{1}})}{|q_I-\mathscr R^i q_{k_{1}}|^3} \right) \\
  \label{eqn:PI}
\end{eqnarray}
We write $q_{I}=(x_I,y_I,z_I)$ and $p_I=(p_{x_I},p_{y_I},p_{z_I})$
and by hypothesis we have $q_I\in\mathcal L$ and $q_I\wedge p_I=0$, it means
that $q_I=(0,0,z_I)$ and $p_I=(0,0,p_{z_I})$.
Finally, we note that $\sum_i \mathscr R^i q_{k_1}=(0,0,r z_{k_1})$,
 obtaining the vector field as
\begin{eqnarray}
  \dot q_I &=& \left(0,0,\frac{1}{m_I}p_{z_I}\right)\\
  \dot p_I &=& \left(0,0,-m_I\sum_{k=1}^s r\hat m_k
            \frac{z_I-z_{k_1}}{|z_I-q_{k_1}|^3}\right),
  \label{eqn:champ:inv}
\end{eqnarray}
which confirms that $\mathcal L$ is invariant under the dynamics of the infinitesimal body.
$\hfill\square$

We assume as known the solution
$\varphi: M\times I\to M$
of the $N$ body problem, defined by
$\varphi(t)=\varphi( \varphi_0;t)$ with
$\varphi_0$ an $\mathscr R$-symmetric initial condition.
Adding a second infinitesimal body with  the same conditions as those of
Proposition (\ref{prop:N+1}), we obtain an $N+2$
body problem such that both infinitesimals have masses of the same order and
they evolve in the vertical line for $t\in(a,b)\subset (\alpha,\beta)$.

Without lost of generality, we can assume that the infinitesimal bodies
have indices
$i=1,2$,
with coordinates $q_i=(x_i,y_i,z_i)$ and $p_i=(p_{x_i},p_{y_i},p_{z_i})$
and any element ${\bf x}\in M$ is written as
\begin{eqnarray}
  {\bf x} &=& ( (x_1,y_1,z_1),\cdots,(x_{m},y_{m},z_{m}),
    (p_{x_1},p_{y_1},p_{z_1}),\cdots,(p_{x_{m}},p_{y_{m}},p_{z_{m}}))^T,
  \label{eqn:point}
\end{eqnarray}
where $m=N+2$.

Proposition \ref{prop:N+1} permits us to denote
the position and momenta of each
infinitesimal
by $(q_{i},p_{i}) = (0,0,z_i,0,0,p_{z_i})$, $i=1,2$,
and their masses by $m_{1}$ and $m_{2}$,
the dynamics of both infinitesimals is given by the Hamiltonian vector field
\begin{eqnarray}
  \dot z_{1} = \frac{1}{m_{1}}p_{z_1},\quad
  \dot p_{z_1} = -m_{1}\sum_{k=1}^s
    r\hat m_k \frac{z_1-z_{k}}{|z_1-q_{k_1}|^3}
    -\frac{m_{1}m_{2}}{|z_{1}-z_{2}|^2},\nonumber\\
  \dot z_{2} = \frac{1}{m_{2}}p_{z_2},\quad
  \dot p_{z_2} = -m_{2}\sum_{k=1}^s
    r\hat m_k \frac{z_2-z_{k}}{|z_2-q_{k_1}|^3}
    +\frac{m_{1}m_{2}}{|z_{1}-z_{2}|^2},
  \label{eqn:champTot}
\end{eqnarray}
with  Hamiltonian function
\begin{eqnarray}
  H(z_{1},z_{2},p_{z_1},p_{z_2},t) &=& \frac{1}{2m_{1}}p_{z_1}^2+
  \frac{1}{2m_{2}}p_{z_2}^2 - m_{1}\sum_{k=1}^s
    \left(
        \sum_{j=1}^r \frac{\hat m_k}{|z_{1}-\mathscr R^{j-1} q_{k_1}|}
    \right) \nonumber\\
  & & - m_{2}\sum_{k=1}^s
    \left(
        \sum_{j=1}^r  \frac{\hat m_k}{|z_{2}-\mathscr R^{j-1} q_{k_1}|}
    \right) - \frac{m_{1}m_{2}}{|z_{1}-z_{2}|}
  \label{eqn:HamNew}
\end{eqnarray}
where $q_{k_j}=\mathscr R^{j-1}q_{k_1}(t)$, $j=1,\cdots,r$. Elements $q_{k_1}$, for $k=1,\cdots,s$,
are representatives of every cyclic subset under $\mathscr R$  for which $|z_i-q_{k_1}|=
|z_i-\mathscr R ^j q_{k_1}|$  holds for $i=1,2$, and $j=1,\cdots,r$,  that we have
used in (\ref{eqn:champTot}) to simplify
the expression of the vector field.
\begin{remark}
  It is important to note that the evolution of the $N$ primary
  bodies is not a relative equilibrium in general; this is the case of the
  Hip-Hop and other more general solutions. As a consequence, it is not always
  possible to reduce the dimension of the vector field. However, some
  configurations as the $M$-Circular Sitnikov problem
  \cite{MMar1} and the Sitnikov restricted $N$ body problem \cite{Bou1}
  are
  reducible to one degree of freedom
  \footnote{In \cite{MMar1},
  Marchesin defines the $M$-Circular Sitnikov Problem
  where the configuration has $N$
  primary bodies of mass $m$ in circular relative equilibrium
  such that $M=mN$,  and the infinitesimal in the conventional way.
  In \cite{Bou1}, Bountis and Papadakis denote the Sitnikov restricted $N$
  body problem the configuration with $N-1$ primaries in circular relative
  equilibria. Here we  follow the convention  that $N+\nu$  means $N$
  primary and $\nu$ infinitesimal bodies as in \cite{Jim4}.}.
\end{remark}
Every solution of (\ref{eqn:HamNew}) will have a singularity due
to collision when $|z_{1}-z_{2}|\to 0$ at $t\to b$ if
$b\in (\alpha,\beta)$.
We can regularize this type of singularities in order to extend
the solutions of (\ref{eqn:HamNew}) for every $t\in(\alpha,\beta)$.
Before stating the  main result of this section, we prove some technical lemmas
which simplify computations of the regularizing transformation.

\begin{lem}\label{lem:B}
  The linear transformation $T_B \in Aut(M)$ with associated matrix 
  $B\in M_{m\times m}(\mathbb R)$, with $m=6(N+2)$,
  which sends
\begin{eqnarray*}
     (z_1,z_2)&\mapsto&
               \left(z_1-z_2, (1-\mu) z_1+\mu z_2\right), \qquad \mu\in(0,1/2]\\
           (p_{z_1}, p_{z_2})&\mapsto&
           \left( \mu p_{z_1} - (1-\mu) p_{z_2}, p_{z_1}+p_{z_2} \right),
\end{eqnarray*}
and fixes all other components, is symplectic.
\end{lem}
{\it Proof.}
It is sufficient to show that the reduced matrix
\begin{eqnarray}
  \hat B = \left( \begin{array}[]{cccc}
     1 & -1 & 0 & 0\\
     (1-\mu) & \mu& 0 & 0\\
      0 & 0& \mu & -(1-\mu)\\
     0&0&1&1
  \end{array}\right)
  \label{eqn:mat:A}
\end{eqnarray}
is symplectic and a straightforward computation shows that indeed 
$\hat{B}^TJ\hat{B}=J$.
$\hfill\square$

Since the masses of the infinitesimal bodies have the same order, it
is possible to  write a linear relation in the form
$m_{2}=c\ m_{1}$ for some constant $c\in (0,1]$. Then the Hamiltonian
function and vector field can be written in a more symmetric way.

\begin{lem}\label{lem:Scal}
  The parameters $m$ and $\epsilon$ defined by
\begin{eqnarray}
  m=\frac{m_{1}+m_{2}}{2},\qquad {\rm and }\qquad
  \epsilon=\frac{m_{1}-m_{2}}{m_{1}+m_{2}},\label{eqn:par}
\end{eqnarray}
and the time rescaling
\begin{eqnarray}
   \hat t= mt
\end{eqnarray}
take $X_H$ and $H$ defined in (\ref{eqn:champTot}) and (\ref{eqn:HamNew})
respectively to the form
\begin{eqnarray}
   z_{1}^\prime = \frac{1}{1+\epsilon}p_{z_1},\quad
   p_{z_1}^\prime = -(1+\epsilon)\sum_{k=1}^{s}
    r\hat m_k \frac{z_1-z_{k_1}}{|z_1-q_{k_1}|^3}
    -m\frac{1-\epsilon^2}{|z_{1}-z_{2}|^2},\nonumber\\
   z_{2}^\prime = \frac{1}{1-\epsilon}p_{z_2},\quad
   p_{z_2}^\prime = -(1-\epsilon)\sum_{k=1}^{s}
    r\hat m_k \frac{z_2-z_{k_1}}{|z_2-q_{k_1}|^3}
    +m\frac{1-\epsilon^2}{|z_{1}-z_{2}|^2},
  \label{eqn:champTot:red}
\end{eqnarray}
and
\begin{eqnarray}
  \hat H &=& \frac{1}{2(1+\epsilon)}p_{z_1}^2+
  \frac{1}{2(1-\epsilon)}p_{z_2}^2 - (1+\epsilon)
  \sum_{k=1}^s
    \left(
        \sum_{j=1}^r \frac{\hat m_k}{|z_{1}-\mathscr R^{j-1} q_{k_1}|}
    \right) \nonumber\\
  & & - (1-\epsilon) \sum_{k=1}^s
    \left(
        \sum_{j=1}^r  \frac{\hat m_k}{|z_{2}-\mathscr R^{j-1} q_{k_1}|}
    \right) - m\frac{1-\epsilon^2}{|z_{1}-z_{2}|}
  \label{eqn:HamNew:red}
\end{eqnarray}
where $\hat H =\hat H(z_{1},z_{2},p_{z_1},p_{z_2},\hat t)$, $^\prime=\frac{d}{d\hat t}$ and  $\hat H=\frac{1}{m}H$.
\end{lem}
{\it Proof.}
 By direct substitution of the new parameters (\ref{eqn:par}) into
(\ref{eqn:champTot}) and (\ref{eqn:HamNew}),  we get
expressions (\ref{eqn:champTot:red}) and (\ref{eqn:HamNew:red}).
$\hfill\square$

\begin{theorem}
The binary collisions between the secondary bodies of the
$\mathscr R$-symmetric $N+2$ body problem with Hamiltonian function
(\ref{eqn:HamNew}) are regularizables by the composition of a
linear symplectic transformation  $A\in Sp(M)$ and the Euler regularization
$(i_\xi,dt/d\tau)$ of the rectilinear binary collisions.

\end{theorem}
{\it Proof.} First, we use Lemma \ref{lem:Scal} to work with normalized
infinitesimal masses in such a way that Hamiltonian function
(\ref{eqn:HamNew}) and vector field (\ref{eqn:champTot}) are transformed
to
(\ref{eqn:HamNew:red}) and (\ref{eqn:champTot:red}) respectively which depend
on $\epsilon$ and $m$ as parameters.

Let $M$ be the phase space of the $\mathscr R$-symmetric
$N$+2 body problem such that $M$ is a cone in the total cotanget bundle
$M\subset T^*\mathbb R^{3(N+2)}$.
Since the evolution of the $N$ primaries is under a $\mathscr R$-symmetry,
we can consider that $\mathcal L= \{ {\bf u}\in\mathbb R^3 \ |
\ {\bf u}=(0,0,\tau), \tau\in\mathbb R \}$ is the symmetry axis of $\mathscr R$.

By Lemma  (\ref{prop:N+1}),
$\mathcal L$ is invariant under the evolution of the secondaries, then
 we are concerned with the third component of their coordinates
 $q_{I_i}=(0,0,z_{I_i})$, and $p_{I_i}=(0,0,p_{z_{I_i}})$, for $i=1,2$.
Since the indexing of coordinates will be tediuous we will assume that
$I_1=1$ and $I_2=2$ and the primaries will have coordinates
$q_i=(x_i,y_i,z_i)$ and $p_i=(p_{x_i},p_{y_i},p_{z_i})$, for $3\le i \le N+2$,
which permits to express a single point ${\bf x}\in M$ in the form 
(\ref{eqn:point}).
We select $M$ to be the cone which holds $z_1>z_2$ such that the infinitesimal
masses are in relation $m_1\ge m_2$.

Consider the transformation $T_E\in Aut(M)$ which
permutes the coordinates with indices
\begin{eqnarray}
  \fl
   \left(
   \begin{array}[h]{cccccccccccccc}
     1& 2 & 3 &4 & 5 & 6 & 7 & \cdots &i&\cdots & n+3 & n+4&n+5& n+6\\
     5 & 6 & 1 & 7 & 8 & 3 &9 &\cdots& i+2 &\cdots&2 & n+5 & n+6  & 4
   \end{array}
      \right),
  \label{eqn:perm}
\end{eqnarray}
where $n={\rm dim}(M)/2$,
in such a way that the determinant of the associated matrix $E$ is unity.

Matrix $E$ have the following properties
\begin{itemize}
  \item $E^T=E^{-1}$,
  \item $E^TJE=J$.
\end{itemize}
Define $A=EB$ where $B$ is the associated matrix of transformation
from Lemma \ref{lem:B} and $A\in Sp(M)$ since $E,B\in Sp(M)$.
Then $A$ sends any element ${\bf x}\in M$ to
\begin{eqnarray}
  A\cdot {\bf x}
  &=& (z_{1}-z_{2},\mu p_{z_{1}}-(1-\mu) p_{z_{2}}, x_{i_1},\cdots , p_{i_j} )^T,
  \quad \mu\in \left(0,1/2 \right],\label{eqn:V}
\end{eqnarray}
where $\mu=\frac{1-\epsilon}{2}$ and $(i_1,\cdots,i_j)$ is an even permutation of the other $6(N+2)-4 $ indices.

The first two components of (\ref{eqn:V}) define a symplectic subspace
$(\mathcal V,\omega|_{\mathcal V})$ of the phase space $(M,\omega)$.
The other elements, define another symplectic subspace
$(\mathcal V^\omega,\omega|_{\mathcal V^\omega})$ which is $\omega$-orthogonal
such that $M=\mathcal V\oplus \mathcal V^\omega$.
Singularities due to binary collision of the secondary bodies belong to the subspace
$(\mathcal V,\omega|_{\mathcal V})$ and this  happens when the first component goes to
zero.

Now, we can apply the canonical inclusion of the Euler transformation
to  $A$ in the form
\begin{eqnarray*}
  {\bf z}&=& ( i_\xi \circ A ){\bf x}, \qquad {\bf z},{\bf x}\in M,
\end{eqnarray*}
in such
a way that the transformation
\begin{eqnarray}
  \rho &:=& (i_\xi\circ A)^{-1}
\end{eqnarray}
and the rescaling time $dt=2\mu(1-\mu)Q_1^2d\tau$ regularizes analytically
the binary collisions of the system (\ref{eqn:champTot:red}) and
(\ref{eqn:HamNew:red})  .
Since the set ${\rm Sp}_{(x,\xi)}(M,\omega)$ is a group under composition, we have immediately that
transformation (\ref{eqn:coord}) is also symplectic on the
manifold $( M,\omega)$ and the following diagram commutes
\begin{eqnarray*}
\xymatrix{
   M\ar[r]^{i_\xi} \ar[drr]_{H\circ \rho}
   & M \ar[dr]^{H\circ A} \ar[r]^{A}& M
   \ar@/_2pc/[ll]_{\rho}
   \ar[d]^{H} \\
    &  &  \mathbb R    }
\end{eqnarray*}

The regularized phase space will have the form
\begin{eqnarray}
   {i_\xi} :\mathcal V \oplus \mathcal V^\omega \rightarrow
      \mathcal R \oplus \mathcal V^\omega.
\end{eqnarray}
where $\mathcal R $ is the symplectic subspace $\mathcal V$ transformed
under the Euler regularization.

Denoting ${\bf z}=(Q_1,Q_2,\cdots, Q_n,P_1,P_2,\cdots,P_n)\in M$
we get in local coordinates the expression
for ${\bf x}= \rho({\bf z})$ given by
\begin{eqnarray}
 z_1 = Q_2 + (1-\mu)\frac{Q_1^2}{2},&\qquad& p_{z_1}= \mu P_2 +
 \frac{P_1}{Q_1},\label{eqn:z1}\\
 z_2 = Q_2 - \mu\frac{Q_1^2}{2},&\qquad& p_{z_2}= (1-\mu) P_2 -
 \frac{P_1}{Q_1}.\label{eqn:z2}
  \label{eqn:ch:gral}
\end{eqnarray}
All other components obey the rules of indexing of ${\bf x}$ given in
(\ref{eqn:point}) and permutation (\ref{eqn:perm}).
Substituting (\ref{eqn:z1}) and (\ref{eqn:z2}) into (\ref{eqn:champTot:red})
and (\ref{eqn:HamNew:red}) and applying the time rescaling with
$\mu=\frac{1-\epsilon}{2}$ and  $(1-\mu)=\frac{1+\epsilon}{2}$
we obtain the
Hamiltonian function in the form

\begin{eqnarray}
\fl
\Gamma &=& \frac{1}{2} \left( \frac{1-\epsilon^2}{4} P_2^2 Q_1^2 + P_1^2\right)
- \frac{1-\epsilon^2}{2} Q_1^2\left[ V_1(Q_1,Q_2) + h  \right] - (1-\epsilon^2)^2 m .\label{eqn:Gamma}
\end{eqnarray}
where
\begin{eqnarray*}
V_1({\bf Q}) &=& \sum_{k=1}^s
    \sum_{j=1}^r
    \left(
    \frac{(1+\epsilon)\hat m_k}{|Q_2+\frac{1+\epsilon}{4}Q_1^2 -Q_{k_j}|}
    - \frac{(1-\epsilon) \hat m_k}{Q_2-\frac{1-\epsilon}{4}Q_1^2- Q_{k_j}|}
    \right),\\
     \qquad{\bf Q}&=&( (Q_1,Q_2,0,0,0,0),(Q_{1_1},\cdots,Q_{s_r}) ),
\end{eqnarray*}
and $Q_{k_j}=q_{k_j}$, $k=1,\cdots,s$, $j=1,\cdots,r$, are the positions of the $N$ primaries.
Since $\rho$ is locally symplectic
on the open manifold with boundary $(M,\omega)$
the new Hamiltonian vector field
can be obtained directly from the regularized Hamiltonian function
\begin{eqnarray}
\Gamma:&M\times \mathbb R &\to \mathbb R\\
&({\bf z},h) &\mapsto \Gamma( {\bf z};h)
\end{eqnarray}
which depends on the parameter
$h\in\mathbb R$.
The regularized Hamiltonian vector field $X_{\Gamma_h}$ which also depends
on the energy level $h$ has the form
\begin{eqnarray}
\begin{array}{ccl}
    Q_1^\prime &=&  P_1,\\
    Q_2^\prime &=& \frac{1-\epsilon^2}{4} Q_1^2P_2,\\
    P_1^\prime &=& -\frac{1-\epsilon^2}{4}Q_1\left( P_2^2 -
    4  \left[ V_1({\bf Q}) + h \right]
    -2Q_1 \cdot \frac{\partial V_1}{\partial Q_1}\right), \\
    P_2^\prime &=& -\frac{1-\epsilon^2}{2} Q_1^2 \cdot
    \frac{\partial V_1}{\partial Q_2}.
\end{array}
\label{eqn:sig:Gamma}
\end{eqnarray}

Expressions (\ref{eqn:Gamma}) and (\ref{eqn:sig:Gamma}) are free of
singularities due to collisions between the secondary bodies. In this way,
we have extended the Hamiltonian system to the set $\Delta:=\{z_1=z_2\}$
which is a subset of the boundary $\partial M$.

\hfill$\square$

\begin{remark}
The energy levels $H({\bf x})=h$ are mapped to the zero set
of $\Gamma$ and we will denote them by
\begin{eqnarray}
  \Sigma_h =\left\{ {\bf z}\in M \ |\ \Gamma({\bf z};h)=0, \ h\in{\rm Img}(H)
    \subset \mathbb R \right\}
  \label{set:level}
\end{eqnarray}
regardless of wheather this is the energy level in the original system or
in the regularized one.

The Hamiltonian vector field $X_{\Gamma_h}$ is valid only on the
energy level $\Sigma_h$ for every $h\in {\rm Img}(H)$ fixed.
\end{remark}

\begin{figure}
 \centering
 \includegraphics[scale=0.5]{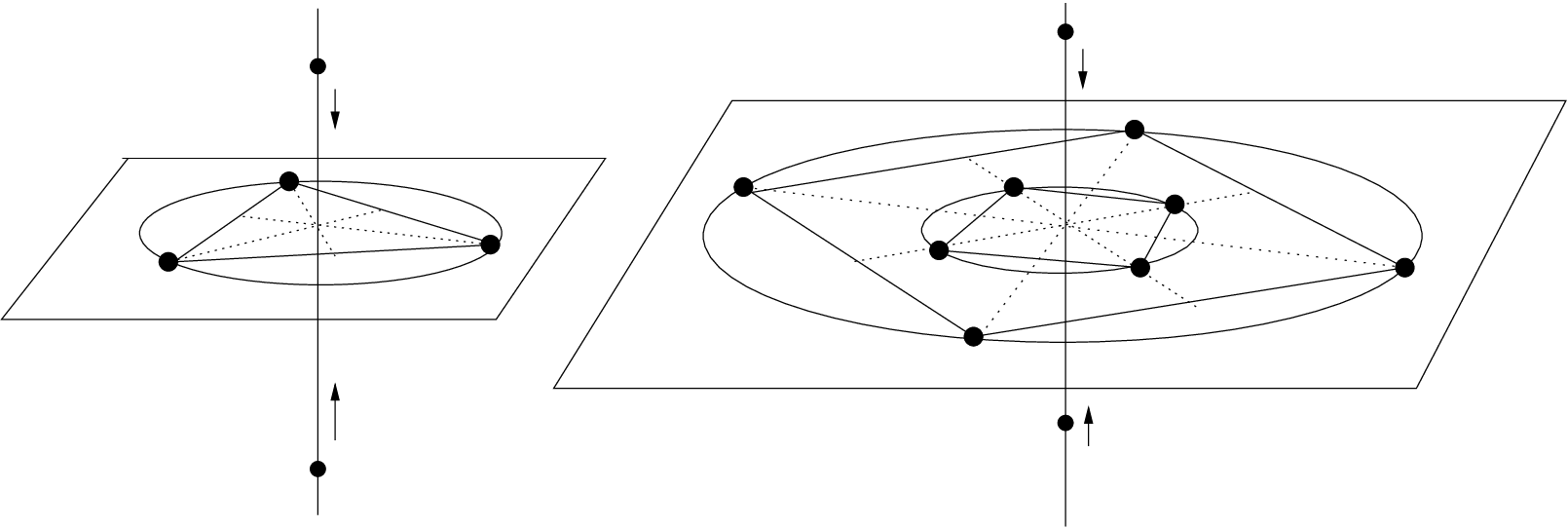}
 \caption{Some circular collinear $N+2$ problems.\label{fig:other}}
\end{figure}

\begin{figure}
 \centering
 \includegraphics[scale=0.4]{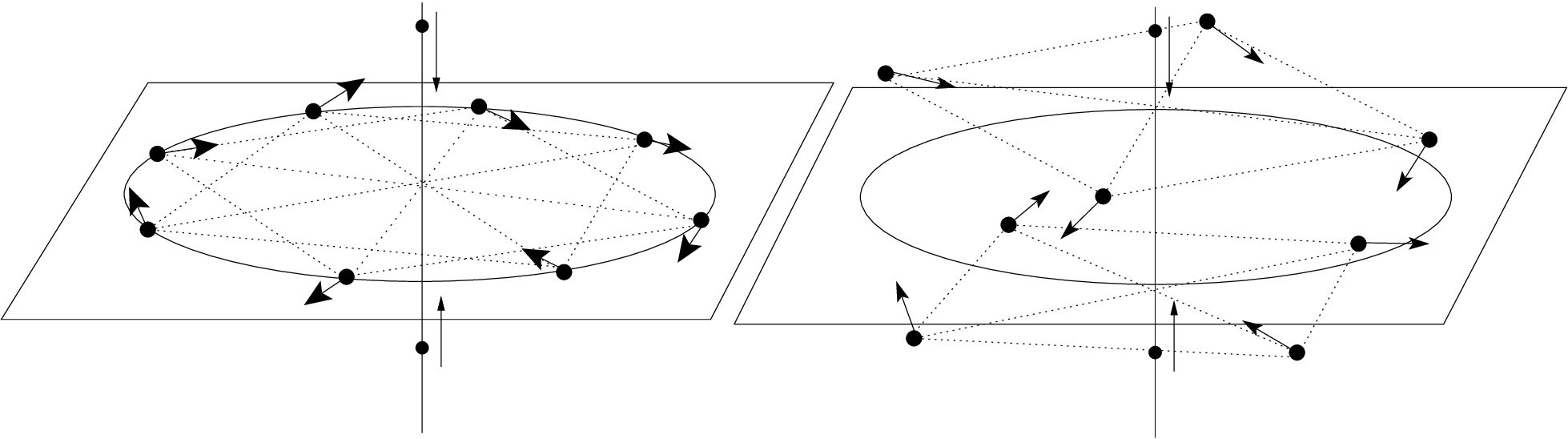}
 \caption{HipHop-collinear $2N+2$ problem.\label{fig:hiphop}}
\end{figure}

Examples of this type of systems are the circular collinear $N+2$ and
$2N+2$ problems shown in Figure \ref{fig:other}. Other examples are
constructed with $2N$ massive bodies in a Hip-Hop solution and 2
infinitesimals bodies on the line determined by the angular moment
of the system as the reader can see in Figure \ref{fig:hiphop}.

Now, we proceed to study a special case of the $N$+2 body problem with many
symmetries. We called this problem the circular $N$+2 Sitnikov problem \cite{Jim2}
since this is a generalization of the circular $N$ Sitnikov problem
\cite{MMar1, Bou1},  obtained by adding another infinitesimal body.

\section{The Circular N+2 Sitnikov Problem\label{chap:FBSP}}

For an application of the symplectic regularization, we select
a special configuration of the restricted $N+2$ body problem.
This particular configuration
has $N$ massive bodies
with masses $m_3=\cdots=m_{N+2}=\frac{1}{N}$ in relative
equilibrium evolving in circular orbits on the vertices of
a regular $N$-gon
around their center of masses. 
The system has two infinitesimal bodies that evolve  on the perpendicular
straight line which passes across the center of masses
of the massive bodies.
The massive bodies are called primaries and the infinitesimal bodies
are known as secondaries.
   The problem consists in determining the evolution of the
secondaries under the attraction of primaries
with Newtonian gravitational potential (see Figure 3).
In general, the secondaries have
different infinitesimal masses $m_1\neq m_2$ and without lost of generality
we can assume that $m_2\leq m_1\ll\frac{1}{N}$.

Let $\mathcal Q$ be the configuration space defined by
\begin{eqnarray*}
    \mathcal Q=\{(q_1,q_2)\in\mathbb R^2 | q_1>q_2\},
\end{eqnarray*}
where $q_i$ is the position of the body with mass $m_i$ for $i=1,2$.
The potential function $V:\mathcal Q\to \mathbb R$ of the circular
$N+2$ Sitnikov problem is
\begin{eqnarray*}
  V(q_1,q_2)&=&  \frac{m_1}{\sqrt{q_1^2+r^2}} + \frac{m_2}{\sqrt{q_2^2+r^2}}  +
     \frac{m_1m_2}{q_1-q_2},
\end{eqnarray*}
and the Hamiltonian function $H:T^*\mathcal Q\to \mathbb R$ will be
\begin{eqnarray}
   H({\bf q},{\bf p}) &=& \frac{1}{2} {\bf p}^T \mathcal M^{-1} {\bf p} - V({\bf q}),\label{eqn:ham02}
\end{eqnarray}
where ${\bf q}=(q_1,q_2)$ is the vector of positions, ${\bf p}=(p_1,p_2)$ is the vector of conjugate momenta and
$\mathcal M={\rm diag}\left(m_1,m_2
\right)$ is the matrix of masses. The constant of universal gravitation
is $G=1$ and $r$ is the radius of the  circle
which contains the vertices of the $N$-gon and must fulfill the following conditions
\cite{Lin2}
\begin{eqnarray*}
 2w^2r^3 =Gm\sum_{\gamma=1}^\nu\frac{1}{\sin(\frac{\pi\gamma}{N})},
    \qquad\qquad N=2\nu+1,\quad,\nu\in\mathbb N\\
    2w^2r^3 =Gm \left(\frac{1}{2}+\sum_{\gamma=1}^{\nu-1}\frac{1}
    {\sin(\frac{\pi\gamma}{N})} \right),\quad N=2\nu,
\end{eqnarray*}
where $w$ is the angular velocity and $m$ is the mass of each primary in the
circular relative equilibrium. Since $G=1$ and $m=\frac{1}{N}$, and
considering $w=1$ we obtain the relation
\begin{eqnarray*}
  r^3 = \frac{1}{2N} \sum_{\gamma=1}^\nu\frac{1}{ \sin(\frac{\pi\gamma}{N})}\quad
  {\rm or}\quad
  r^3 = \frac{1}{2N} \left( \frac{1}{2}+\sum_{\gamma=1}^{\nu-1}\frac{1}{ \sin(\frac{\pi\gamma}{N})}\right),
\end{eqnarray*}
whenever $N$ is odd or even respectively.

 Corresponding expressions  were found by
Bountis and Papadakis in \cite{Bou1} in the $N$+1 Sitnikov problem
where the value for $r$ is given by
\begin{eqnarray*}
  r= \frac{1}{2}\csc \left( \frac{\pi}{N} \right),
\end{eqnarray*}
and the masses of the primaries are $m=\frac{1}{K}$ with
\begin{eqnarray*}
  K = \sqrt{2(1-\cos 2\theta)}\sum_{i=2}^N \frac{\sin^2\theta\cos(\frac{\nu}{2}+1-i)\theta}{\sin^2 \left( \nu+1-i \right)\theta}.
\end{eqnarray*}
Marchesin \cite{MMar1} in contrast, fixes the radius $r=\frac{1}{2}$ and by
a suitable rescaling studies the effect of the variation of primary masses
on the period function $T(h)=T(h;m)$. In general, a suitable change on the
angular velocity and the masses of the primaries allows us to normalize the
radius to $r=1$ (which is not the case in this paper).

Applying Lemma \ref{lem:Scal} we will write the masses of the secondary bodies
as $m_1=m(1+\epsilon) $ and $m_2=m(1-\epsilon)$ and the time rescaling
$t\mapsto mt$ will produce the reduced masses $\alpha=1+\epsilon$ and
$\beta=1-\epsilon$.
Denoting by $r=r_N$ the radius of the  circle for the
$N$+2 Sitnikov problem,
the potential function now depends on the number of primary bodies as
a parameter, then $V:\mathcal Q\times \mathbb N\to \mathbb R$
becomes
\begin{eqnarray*}
  V(q_1,q_2;N)&=&  \frac{1+\epsilon}{\sqrt{q_1^2+r_N^2}} + \frac{1-\epsilon}{\sqrt{q_2^2+r_N^2}}  +
     m\frac{1-\epsilon^2}{q_1-q_2},
\end{eqnarray*}
 and the Hamiltonian function becomes
\begin{eqnarray}
 \fl
 \quad H({\bf q},{\bf p};N) &=& \frac{1}{2(1+\epsilon)} p_1^2 + \frac{1}{2(1-\epsilon)}
    p_2^2 -\frac{1+\epsilon}{\sqrt{q_1^2+r_N^2}} - \frac{1-\epsilon}
    {\sqrt{q_2^2+r_N^2}} - m\frac{1-\epsilon^2}{q_1-q_2}.\label{fun:Ham}
\end{eqnarray}

It is important to note that the angular velocity of the primary bodies
is  not any more the unity $w\neq 1$ due to the time rescaling $t\to mt$,
however this fact is not relevant
when we restrict the study to the rectilinear (non-perturbed) case.

\begin{remark}
The symmetry $(q_1,p_1,q_2,p_2,\epsilon)\mapsto (q_2,p_2,q_1,p_1,-\epsilon)$ restricts
the analysis to non negative values of the parameter $\epsilon$.
\end{remark}

Let $M=T^*\mathcal Q$ be the phase space of the Hamiltonian system $\mathcal H=(M,\omega,X_H)$
associated to the problem, where $\omega=\sum_{i}^{}dp_i\wedge dq_i$ is the standard symplectic form on $M$.
$\Delta=\{(q_1,q_2,p_1,p_2) \in \mathbb R^4| q_1=q_2\}$ is the set of singularities
of $H({\bf q},{\bf p};N)$ due to collisions and it is easy to see that $\Delta=\partial M$.

The Hamiltonian vector field $X_H$ in local
coordinates is as follows
\begin{eqnarray*}
  \dot q_1 =  \frac{1}{1+\epsilon}p_1,&\hspace{20pt}&   \dot p_1 = -\frac{(1+\epsilon)  q_1}{(q_1^2+r_N^2)^\frac{3}{2}} - m\frac{1-\epsilon^2}{(q_1-q_2)^2},\\
  \dot q_2 =  \frac{1}{1-\epsilon}p_2,&\hspace{20pt}&
  \dot p_2 = -\frac{(1-\epsilon)  q_2}{(q_2^2+r_N^2)^\frac{3}{2}} + m\frac{1-\epsilon^2}{(q_1-q_2)^2}.
\end{eqnarray*}

\begin{figure}
\centering
    \includegraphics[scale=0.4]{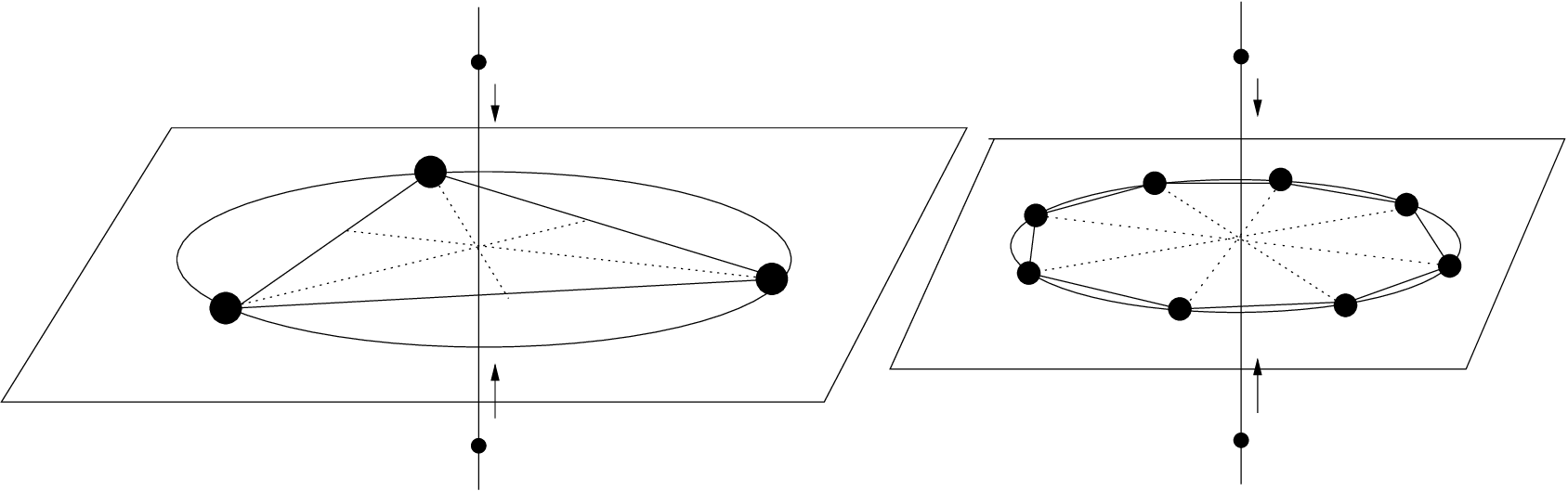}
    \caption{The circular $N+2$ Sitnikov problem for $N=3$ and $N=8$.
    \label{fig:sit}}
\end{figure}
The evolution of both secondaries is restricted to the perpendicular line that
passes by the center of masses of the primaries. The symmetries of the problem
keeps the secondaries on the perpendicular line and since
their angular moment is
null there is not scattering at collisions.

\subsection{Regularization}
To avoid the singularity in both, the Hamiltonian function and the vector field $X_{H}$, we perform a
symplectic regularization. In order  to
extend analytically the equations to the hyperplane $q_1=q_2$ we apply the
transformation $\rho:M\to M$ defined by
\begin{eqnarray}
  q_1 = Q_2+ \frac{1-\epsilon}{4}Q_1^2,&\hspace{30pt}& p_1 = \frac{1+\epsilon}
    {2} P_2 + \frac{P_1}{Q_1},\nonumber\\
    q_2 = Q_2- \frac{1+\epsilon}{4}Q_1^2,&\hspace{30pt}& p_2 =\frac{1-
    \epsilon}{2} P_2 - \frac{P_1}{Q_1},\label{eqn:coord}
\end{eqnarray}
and the time rescaling
\begin{eqnarray}
  \frac{dt}{d\tau} &=& \frac{1-\epsilon^2}{2} Q_1^2.\label{eqn:newtime}
\end{eqnarray}

If we write ${\bf z}=(Q_1, Q_2,P_1,P_2)$ and $\mu=\frac{1-\epsilon}{2}$,
the regularized Hamiltonian function is
\begin{eqnarray*}
  \fl
  \Gamma = \frac{1}{2} \left( \mu(1-\mu) P_2^2 Q_1^2 + P_1^2\right) - 16\mu^2(1-\mu)^2 m \\
  \fl
   \qquad - 2\mu(1-\mu) Q_1^2\left[ \frac{4(1-\mu)}{\sqrt{\left(2 Q_2+\mu Q_1^2\right)^2+4r_N^2}}
  + \frac{4\mu}{\sqrt{\left(2 Q_2-(1-\mu) Q_1^2\right)^2+4r_N^2}} +h \right].\nonumber
\end{eqnarray*}
We denote $\Gamma_h({\bf z},\mu;\epsilon)=\Gamma({\bf z},\mu;\epsilon,h)$,
and we call  the triplet $(\hat M, \omega, X_{\Gamma_h({\bf z},\mu)})$ the regularized
system, where $\hat M= T^*(\mathcal Q\cup \{q_1=q_2\})$ and
 $ X_{ \Gamma_h}$ is the regularized Hamiltonian
field
\begin{eqnarray}
   \dot {\bf Q} =\frac{\partial \Gamma_h}{\partial {\bf P}},
   &\hspace{30pt}&\dot {\bf P} =-\frac{\partial \Gamma_h}{\partial {\bf Q}}.
\end{eqnarray}
In local coordinates we get
\begin{eqnarray}
\fl
\qquad
\begin{array}{ccl}
    Q_1^\prime &=&  P_1,\\
    Q_2^\prime &=& \mu(1-\mu) Q_1^2P_2,\\
    P_1^\prime &=& -\mu(1-\mu) Q_1 \left( P_2^2 -
    4\left[ V(Q_1,Q_2) + h \right] -
    2Q_1\cdot\frac{\partial V}{\partial Q_1} \right),\\
    P_2^\prime &=& -2\mu(1-\mu) Q_1^2 \cdot \frac{\partial V}{\partial Q_2}.
\end{array}
\end{eqnarray}

Computing the partial derivatives we obtain
\begin{eqnarray}
  \fl
  \qquad
\begin{array}{l}
  \frac{\partial V}{\partial Q_1} = 8\mu(1-\mu)Q_1
          \left(
        \frac{2Q_2-(1-\mu) Q_1^2}{[(2Q_2-(1-\mu) Q_1^2)^2+4r_N^2]^{\frac{3}{2}}}
            - \frac{2Q_2+\mu Q_1^2}{[(2Q_2+\mu Q_1^2)^2+4r_N^2]^{\frac{3}{2}}}
      \right),\\
  \frac{\partial V}{\partial Q_2} =
          -8\left(
        \frac{\mu(2Q_2-(1-\mu) Q_1^2)}{[(2Q_2-(1-\mu) Q_1^2)^2+4r_N^2]^{\frac{3}{2}}}
            + \frac{(1-\mu)(2Q_2+\mu Q_1^2)}{[(2Q_2+\mu Q_1^2)^2+4r_N^2]^{\frac{3}{2}}}
      \right),
\end{array}
  \label{eqn:partials}
\end{eqnarray}
and arranging equivalent terms in the expression
\begin{eqnarray*}
  \fl
\qquad\begin{array}{l}
  4V+2Q_1\cdot \frac{\partial V}{\partial Q_1}= 16\left[
  \frac{\mu(4Q_2^2 - 2(1-\mu)Q_2Q_1^2+ 4r_N^2)}{[(2Q_2-(1-\mu) Q_1^2)^2+4r_N^2]^{\frac{3}{2}}}
  +\frac{(1-\mu)(4Q_2^2 + 2\mu Q_2Q_1^2+ 4r_N^2)}{[(2Q_2+\mu Q_1^2)^2+4r_N^2]^{\frac{3}{2}}}
  \right],
\end{array}
\end{eqnarray*}
we obtain the vector field as
\begin{eqnarray}
\fl
\qquad
\begin{array}{ccl}
    Q_1^\prime &=&  P_1,\\
    Q_2^\prime &=& \mu(1-\mu) Q_1^2P_2,\\
    P_1^\prime &=& -\mu(1-\mu) Q_1 \left( P_2^2 - 4h
    -16
    \left[
        \frac{(1-\mu)(2(2Q_2+\mu Q_1^2)Q_2+4r_N^2)}{
            [(2Q_2+\mu Q_1^2)^2 +4r^2_N]^{\frac{3}{2}}
        }\right.\right.\\
    & & \qquad\qquad\left.\left.    + \frac{\mu (2(2Q_2-(1-\mu)Q_1^2)Q_2+4r^2_N)}{
            [(2Q_2-(1-\mu) Q_1^2)^2 +4r_N^2]^{\frac{3}{2}}
        }
    \right]
    \right), \\
    P_2^\prime &=& -16\mu(1-\mu) Q_1^2
      \left[
        \frac{\mu (2Q_2-(1-\mu) Q_1^2)}{
            \left[(2Q_2-(1-\mu) Q_1^2)^2 +4r_N^2\right]^{\frac{3}{2}}
        }
        +\frac{(1-\mu)(2Q_2+\mu Q_1^2)}{
            \left[(2Q_2+\mu Q_1^2)^2 +4r_N^2\right]^{\frac{3}{2}}
        }
      \right].
\end{array}
\label{eqn:sys:reg}
\end{eqnarray}

Although the form of the new Hamiltonian function and the vector field are quite complicated, the advantage is
that they are
regular in $\bar M:=M\cup\Delta$.
\subsection{Symmetries.}
The regularized Hamiltonian function has a symmetry
in $P_1$ and $P_2$ that reflects the symmetry with respect to
the fictitious  time $\tau$ in the way
\begin{eqnarray}
   (Q_1,Q_2,P_1,P_2,\tau)\mapsto (Q_1,Q_2,-P_1,-P_2,-\tau).
\end{eqnarray}
It is a generic property of mechanical systems.
The symmetry in the
$Q_1$ variable is fictitious due to the transformation $Q_1^2/2=q_1-q_2$.
Finally, applying the change $Q_2\mapsto -Q_2$ it changes the values of
$(1+\epsilon)\mapsto (1-\epsilon)$ and viceversa.

\begin{theorem}\label{prop:Q4:inv}
  The regularized Hamiltonian system $(M,\omega,X_{\Gamma_h})$
is symmetric
with respect to the hyperplane $Q_2=0$ if
$\epsilon =0$. Moreover,
if $\epsilon =0$, the symplectic plane
$$
\mathcal S_1=\{(Q_1,Q_2,P_1,P_2)\in M| Q_2=P_2=0\}
$$
is invariant under the flow of the regularized Hamiltonian vector field
$X_{\Gamma_h}$.
\end{theorem}
{\it Proof.}
Using the
Hamiltonian function $\Gamma_h$ and substituting
$Q_2\to -Q_2$ it remains invariant if
\begin{eqnarray*}
  \fl
\begin{array}{ccl}
 \frac{1-\mu}{\sqrt{\left(2 Q_2+\mu Q_1^2\right)^2+4r_N^2}}
  + \frac{\mu}{\sqrt{\left(2 Q_2-(1-\mu) Q_1^2\right)^2+4r_N^2}} &=&
 \frac{1-\mu}{\sqrt{\left(2 Q_2-\mu Q_1^2\right)^2+4r_N^2}}
  + \frac{\mu}{\sqrt{\left(2 Q_2+ (1-\mu) Q_1^2\right)^2+4r_N^2}}.
\end{array}
\end{eqnarray*}

This identity has as trivial solution
$\mu=1-\mu$ and this holds if and only if $\epsilon =0$.

In order to prove that $\mathcal S_1$ is an invariant plane under the flow we consider
$Q_2\equiv 0$ for every $\tau\in I\subset \mathbb R$. By hypotesis $\epsilon=0$ and consequently
$\mu=\frac{1}{2}$, then the fourth equation
in (\ref{eqn:sys:reg}) implies $P_2^\prime=0$ and therefore $P_2=constant$.
Additionally, $Q_2^\prime\equiv 0$,
but we know that $\mu(1-\mu) \neq0$ and $Q_1$ is not identically zero.
Then $P_2=0$ and we have the reduced system
\begin{eqnarray}
\left.
\begin{array}{ccl}
    Q_1^\prime =  P_1,&\qquad &
    P_1^\prime =
    -8 Q_1\left( \frac{a^2}{\left(Q_1^4 + a^2\right)^{\frac{3}{2}} }+\frac h8 \right),\\
    Q_2^\prime = 0,&\qquad&
    P_2^\prime = 0,
\end{array}
\right. \label{eqn:sys:red}
\end{eqnarray}
where $a=4r_N$.
Consequently, $\mathcal S_1$ is an invariant plane under the flow $\phi(\tau)$ of the
Hamiltonian vector field $X_{\Gamma_h}$.
$\hfill\square$

It is known that Hamiltonian systems $(M,\omega,X_H)$ which have
invariant symmetry planes can be reduced to systems restricted to
 the invariant plane. In fact, each invariant plane corresponds to
some symplectic subspace and vector fields restricted
to symplectic subspaces can be locally integrable. In this example, the flow
$\phi_H(\tau)$ of the Hamiltonian system restricted to the symplectic
subspace $\mathcal S_1$
is equivalent to have the
 secondaries' relative barycenter $m_1q_1+m_2q_2=0$  at the origin.
\begin{definition}
   We define the \emph{symmetric circular N+2 Sitnikov problem}
to the Hamiltonian system $(M,\omega,X_{H})$
where
$\epsilon = 0$ and the initial conditions are symmetric
\end{definition}
It means that $p_0=p_1(0)=-p_2(0)$ and $q_0=q_1(0)=-q_2(0)$.
 we have the following
\begin{corollary}
   The symmetric circular $N$+2 Sitnikov problem
for $m\sim 0$, is integrable.
\end{corollary}
{\it Proof.}   It is an immediate consequence of Theorem \ref{prop:Q4:inv}. Since
the initial conditions are $q_1(t_0)=-q_2(t_0)$ and $p_1(t_0)=-p_2(t_0)$ and
$\epsilon=0$ then $Q_2(\tau_0)= 0$ and $P_2(\tau_0)= 0$. Additionally,
Proposition \ref{prop:Q4:inv} implies that $\mathcal S_1$ is an invariant
symplectic plane then $Q_2(\tau)\equiv 0$ and $P_2(\tau)\equiv 0$ for all
$\tau\in I\subset \mathbb R$ where $I$ is its domain of definition.

Therefore, the symmetric circular $N$+2 Sitnikov problem is a Hamiltonian
system with one degree of freedom. It has as regularized system
$(\tilde M,\omega, X_{\tilde \Gamma})$
with $\tilde M=T^*\mathbb R_*$ and regularized Hamiltonian function
\begin{eqnarray}
  \tilde \Gamma &=& \frac{1}{2} P_1^2  - 4Q_1^2\left( \frac{1}{\sqrt{Q_1^4+a^2}}
  + \frac h8 \right) -m. \label{eqn:ham:red}
\end{eqnarray}
where $a=4r_N$. This is a first integral for the reduced Hamiltonian system when $\tilde \Gamma=0$

$\hfill\square$

\begin{figure}[h]
\centering
    \includegraphics[scale=0.7]{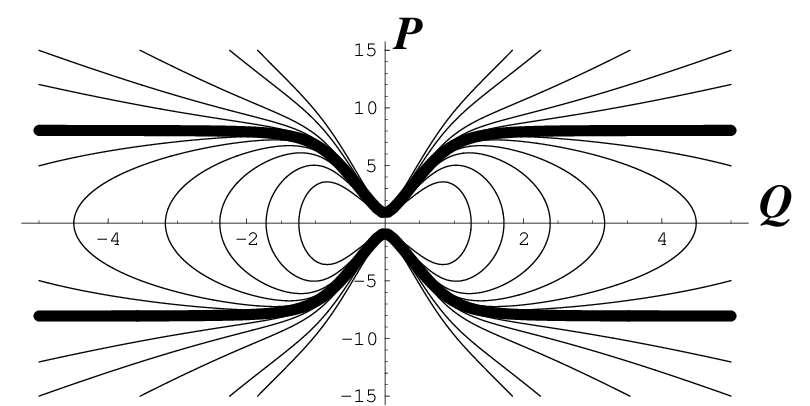}
\caption{Level curves of the symmetric circular $N$+2 Sitnikov problem for $0<m\ll 1/N$.}\label{fig:cur:niv1}
\end{figure}

The vector field $X_{\tilde \Gamma}$ in local coordinates is as  in  the
first  line  in (\ref{eqn:sys:red})
and the level curves are show in Figure \ref{fig:cur:niv1}

\begin{proposition}
   The symmetric circular $N$+2 Sitnikov problem has the following dynamics:
\begin{itemize}
 \item If $h<0$ the solutions are periodic orbits where the secondary bodies collide
   at the origin of coordinates.
 \item If $h=0$ the system has a parabolic solution with the escape of both
   secondaries with null velocity when they reach the infinity.
 \item If $h>0$ the solutions are hyperbolic orbits with escape of both secondaries
   in opposite directions and with positive velocity at infinity.
\end{itemize}

\end{proposition}
{\it Proof.}
We verify this fact directly from the Hamiltonian function of the
original system. Substituting $\epsilon=0$, $q_1(t_0)=-q_2(t_0)$ and
$p_1(t_0)=-p_2(t_0)$ in (\ref{fun:Ham}). Defining  $p:=p_1=-p_2$
 and $q:=q_1=-q_2$ and fixing $H=h$
we obtain
\begin{eqnarray}
  \frac{h}{2} &=& \frac{1}{2}p^2 -\frac{1}{\sqrt{q^2+r_N^2}} -\frac{m}{4q}.\label{red:Ham}
\end{eqnarray}
The maximum distance from the origin that the secondaries can reach is
when $p=0$, then
\begin{eqnarray*}
  \frac{h}{2} &=& -\frac{1}{\sqrt{q^2+r_N^2}} -\frac{m}{4q}.
\end{eqnarray*}
 This has a finite real solution $q>0$ for every fixed $h<0$. It means that the evolution is bounded
and extending the solutions beyond collisions with the regularization, the solutions are periodic
orbits with elastic bouncing at collisions.

On the other hand, if $h\ge 0$ this approach does not apply. For this case, we solve (\ref{red:Ham})
for $p=\dot q$ to obtain
\begin{eqnarray}
    \dot q=\pm\sqrt{ h + \frac{2}{\sqrt{q^2+r_N^2}}+\frac{m}{2q}},
\end{eqnarray}
and we obtain the escape velocity by the limit
\begin{eqnarray*}
    \lim_{q\to\infty} \dot q=\pm\sqrt{h}.
\end{eqnarray*}
Since we  are dealing   with the symmetric problem, we are concerned only with positive values
for $q$, $p$ and $h$. Negative values are associated with the other secondary body.

For $h=0$ the limit $\lim_{q\to\infty} \dot q=0$ implies that the bodies escape
to infinity with  zero velocity, which confirms the parabolic orbit.

Finally, for $h>0$ we have $\lim_{q\to\infty}|\dot q|>0$ and the solutions are
hyperbolic orbits, where the secondary bodies escape to infinity with
positive velocity.

The dynamics is as follows: secondary bodies start its evolution at infinity from
 opposite sides of the plane where the primary bodies evolve. Secondaries approach  the
massive system symmetrically to collide at the origin with ellastic bouncing and escape
to infinity in opposite directions.

$\hfill\square$

\section{Numerical test}
\begin{figure}[h]
  \begin{center}
    \includegraphics[scale=0.59]{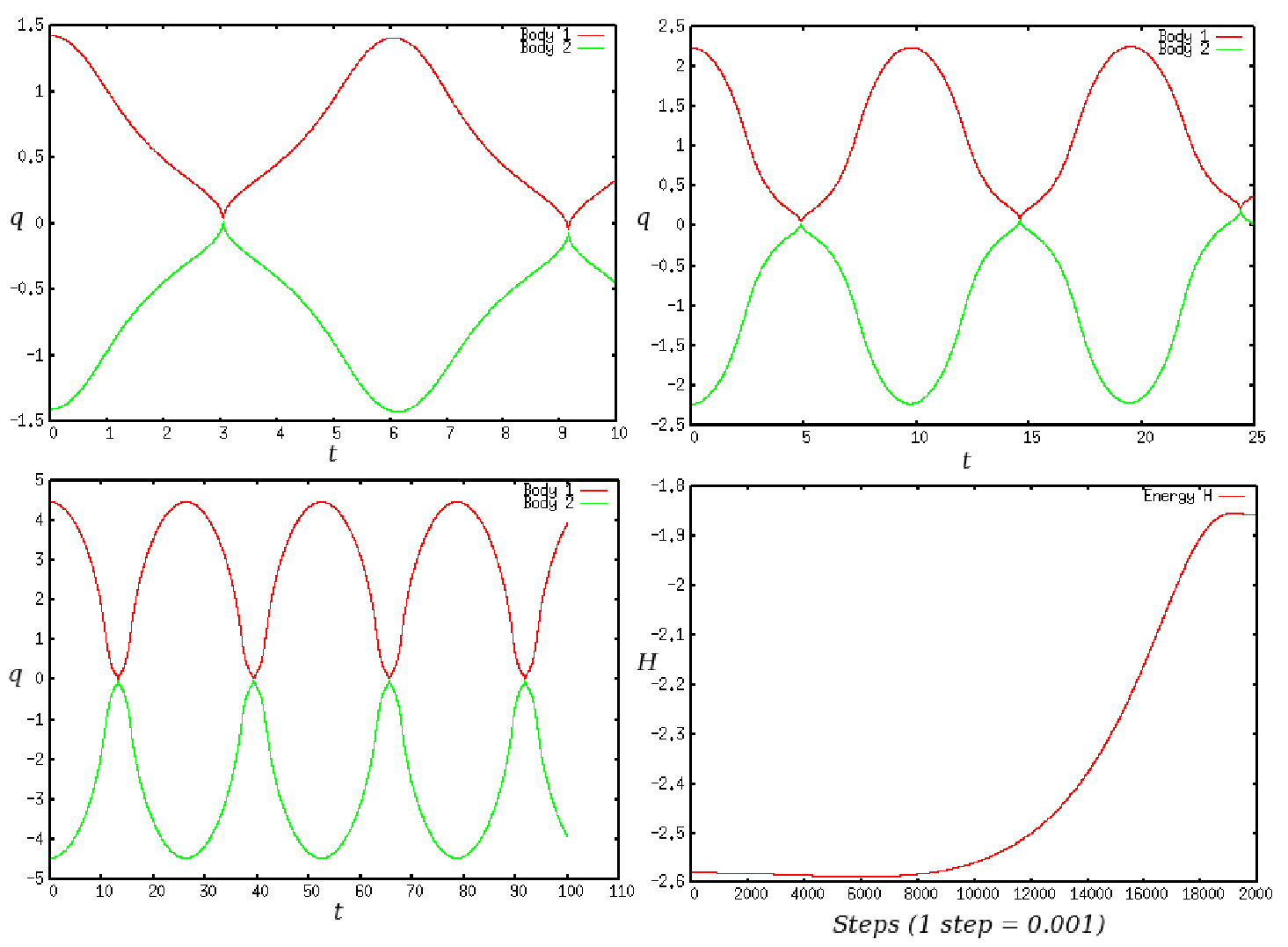}
  \end{center}
  \caption{Some numerical integrations of the regularized system
  with almost symmetric initial conditions: 
  $(Q_1=1,Q_2=0.01,P_1=0,P_2=0)$ 
  (up-left), 
  $(Q_1=5,Q_2=0.01,P_1=0,P_2=0)$ 
  (up-right) and 
  $(Q_1=20,Q_2=0,P_1=0,P_2=0)$ 
  (down-left). The value
  of the mass is $m=10^{-5}$. For mid-term computations, the error in total 
  energy grows quadratically suspected by the quadratic rescaling
  function and the non-separability of the regularized system (down-right).
  }
  \label{fig:graph}
\end{figure}

We have tested the regularized system for the case $m=m_3=m_4$ with values 
$m\in\{10^{-5}, 10^{-7}, 10^{-10}\}$
and almost symmetric initial conditions, which are close to the integrable
symmetric problem.  
We have used a fourth order symplectic integrator of type $\mathcal{SBAB}_2$
with coefficients $(1/6,1/2,2/3,1/2,1/6)$ and timestep 
$\tau=10^{-3}$ (see \cite{Las1} for details about this integrator).
The simulations were programmed in TRIP \cite{TRIP} in double precision. Figure \ref{fig:graph} shows
three test with
$(Q_1,Q_2,P_1,P_2)=(1,10^{-2},0,0)$, 
$(Q_1,Q_2,P_1,P_2)=(5,10^{-2},0,0)$ and
$(Q_1,Q_2,P_1,P_2)=(20,0,0,0)$. 
The value of the energy $H=h$ for a mid-term computation shows a 
non-linear growth (Figure \ref{fig:graph} right below), maybe due 
to the non separability of the regularized system. 
Other factors to this behavior can be the quadratic rescaling function 
$g({\bf Q})=\mu(1-\mu)Q_1^2$ or the size 
of  the ``infinitesimal'' mass $m$.

Finally, in the case $m_3\neq m_4$ the system will experiment
momentum transfer and we need an additional  transition mapping to 
continue the solutions 
beyond collisions \cite{Jim3}. We will perform a complete study of the 
numerical simulations for both cases in a future work.

\section*{Awknowledgments}
We would like to thank the referees for their careful review, the valuable
comments and the references \cite{Mik1,Mik2,Pre1} about \emph{algorithmic 
regularization}. The first author is greatful to  Profr. Laskar and the 
IMCCE for the facilities to perform the numerical computations.

\end{document}